# Secure Communications using Nonlinear Silicon Photonic Keys

**BRIAN C. GRUBEL, BRYAN T. BOSWORTH, MICHAEL R. KOSSEY, A. BRINTON COOPER, MARK A. FOSTER, AND AMY C. FOSTER**[*]

[1]*Department of Electrical and Computer Engineering, Johns Hopkins University, Baltimore, Maryland 21218, USA*
*\*amy.foster@jhu.edu*

**Abstract:** We present a secure communication system constructed using pairs of nonlinear photonic physical unclonable functions (PUFs) that harness physical chaos in integrated silicon micro-cavities. Compared to a large, electronically stored one-time pad, our method provisions large amounts of information within the intrinsically complex nanostructure of the micro-cavities. By probing a micro-cavity with a rapid sequence of spectrally-encoded ultrafast optical pulses and measuring the lightwave responses, we experimentally demonstrate the ability to extract 2.4 Gb of key material from a single micro-cavity device. Subsequently, in a secure communications experiment with pairs of devices, we achieve bit error rates below $10^{-5}$ at code rates of up to 0.1. The PUFs' responses are never transmitted over the channel or stored in digital memory, thus enhancing security of the system. Additionally, the micro-cavity PUFs are extremely small, inexpensive, robust, and fully compatible with telecommunications infrastructure, components, and electronic fabrication. This approach can serve one-time pad or public key exchange applications where high security is required.

## 1. Introduction

Maintaining confidentiality while communicating in the presence of adversaries forms the foundation of cryptology [1]. While modern ciphers attempt to provide such guarantees, the one-time pad (OTP) protocol provides an information-theoretic secure approach even against computationally unbounded adversaries [2]. In practice, the price of such security is storing large digital keys in nonvolatile memory, thus increasing the risk of compromise through duplication of this presumed private information. Additionally, modern systems convey such increasingly vast volumes of data that the storage of suitably long keys is an immense challenge.

Promising recent work on the storage of long OTP keys [3] based on spatial optical scattering within a complex material [4–6] addresses many of the weaknesses of electronic storage, as such devices are difficult to probe, modify, or clone. Such *optical scattering PUFs (OSPUFs)* [7] provide an intriguing alternative to digital storage of private key information. However, OSPUFs are large and difficult to integrate into electronic circuits [6]. Further, mechanical positioning variability introduces inter-key noise that has not been fully mitigated by error correction coding, resulting in a system bit error rate (BER) of $1.7 \times 10^{-1}$ at a highly inefficient code rate of only 0.035 [3,5].

Here we demonstrate an information-theoretically secure symmetric block cipher based on the fuzzy extraction of key material from information-dense ultrafast nonlinear optical interactions in silicon photonic micro-cavities [8,9] [Fig. 1(a)]. The photonic micro-cavities are operated in a nonlinear optical regime and designed as a disk with a chamfer, which exhibits reverberant ray-chaotic behavior [10]. In our previous work [10], we examined the device's Lyapunov exponent, phase space, and escape rate which indicated that the devices operate in a chaotic state. Thus, the cavity's ultrafast optical output waveforms are complex and highly sensitive to the precise physical cavity structure. This high behavioral sensitivity to precise structure prevents an adversary from cloning a micro-cavity. Still, despite this sensitivity, the responses from a given cavity are deterministic and highly repeatable. This combination of robustness and unclonability allows these micro-cavities to serve as a robust PUF [9]. Here we demonstrate the ability to extract long binary keys (2.4 Gb) from a single micro-cavity PUF that can subsequently be used to encrypt a communications channel. Furthermore, in an encrypted communications scenario employing unique pairs of PUFs we demonstrate a BER of $10^{-5}$ at a code rate of 0.1. Notably, at this code rate, our attempts at cloning the PUFs results in a probability of correct message decryption of only $10^{-11}$.

## 2. System Design

### Device Fabrication and Information Extraction

To investigate this approach, we fabricated six unique silicon photonic micro-cavity PUF designs (variations on a 30-µm diameter silicon disk with a chamfer) and two exact copies (clones) of each PUF using electron beam lithography and standard nanofabrication techniques [Fig. 1(a)].

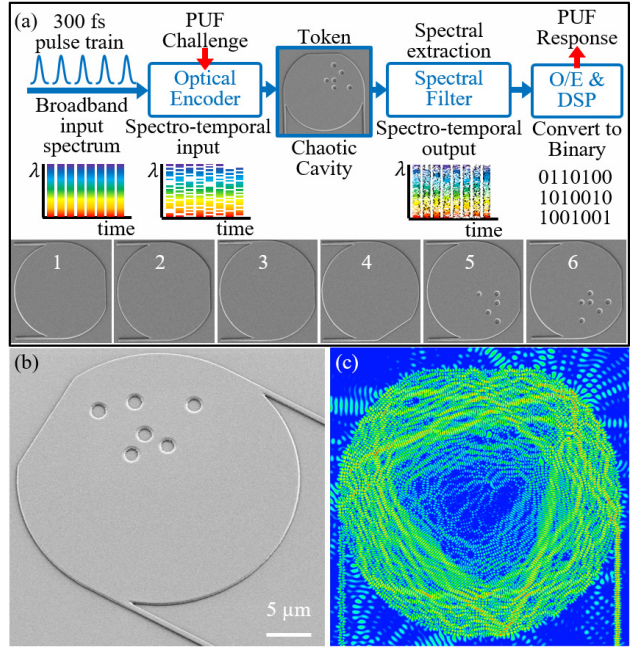

Fig. 1. Operation of the photonic PUF. (a) A spectro-temporally encoded ultrafast optical pulse sequence is sent into the photonic PUF. The response sequence generated by the cavity is passed through a programmable spectral filter and is then detected. A series of digital signal processing steps are performed before sending the binary response to a fuzzy extractor to generate a binary key. Scanning electron microscope (SEM) images of the six prototype designs are shown. (b) Larger SEM image of an example prototype device. (c) A finite difference time domain (FDTD) image detailing the excitation of many transient optical modes within the cavity after ultrafast excitation.

These devices are fabricated from single-crystal silicon-on-insulator (SOI) material (220-nm thick) clad with 1 µm of silicon dioxide, achieving a device volume of approximately 160 µm$^3$. Each design differs from the others in a single parameter including existence, size, and position of the chamfer, as well as the presence of arbitrarily positioned holes within the cavity. In addition, two copies of every cavity are fabricated on the same SOI die and in the same fabrication run, permitting analysis of PUF clonability. All clones use an identical design and are fabricated at the same time and in close proximity to the legitimate devices to maximize the chance of successful cloning. The precise fabrication process is detailed in [9]. Design 1 is 30-µm in diameter and has a 0.8$r$ sized chamfer (0.8 times its radius) centered at 0 degrees on the unit circle; design 2 is 26-µm in diameter and has a 0.8$r$ sized chamfer at 0 degrees; design 3 is 30-µm in diameter and has a 0.7$r$ sized chamfer at 0 degrees; design 4 is

30-µm in diameter and has a 0.8*r* sized chamfer at -45 degrees; and both design 5 and 6 are 30-µm in diameter with 0.8*r* sized chamfers at 0 degrees, each with a different hole pattern. Due to damage from device handling, we were able to test four devices each of designs 1-4, three design 5 devices, and two design 6 devices, yielding 21 experimental devices in total. All designs employ single-mode silicon waveguide ports for robust input and output coupling to the cavity.

To extract a binary key from a photonic micro-cavity PUF, we probe it with a rapid sequence of spectrally amplitude encoded, [11] ultrashort optical pulses [Fig. 1(a)] exciting a unique combination of spatial optical modes that interact with the fine-scale structure of the cavity interior and with one another via the optical nonlinearity of silicon [11,12]. This produces a sequence of ultrafast optical responses, each of which ideally contains independent, information-carrying spectro-temporal features that are sensitive to the cavity structure. To extract information from these responses, we pass each response through a spectral amplitude mask and measure the transmitted pulse energy with a lower bandwidth photodiode. A binary response sequence is then computed from the measured analog pulse energy sequence using a post-processing algorithm (see the following section).

*Experimental Setup*

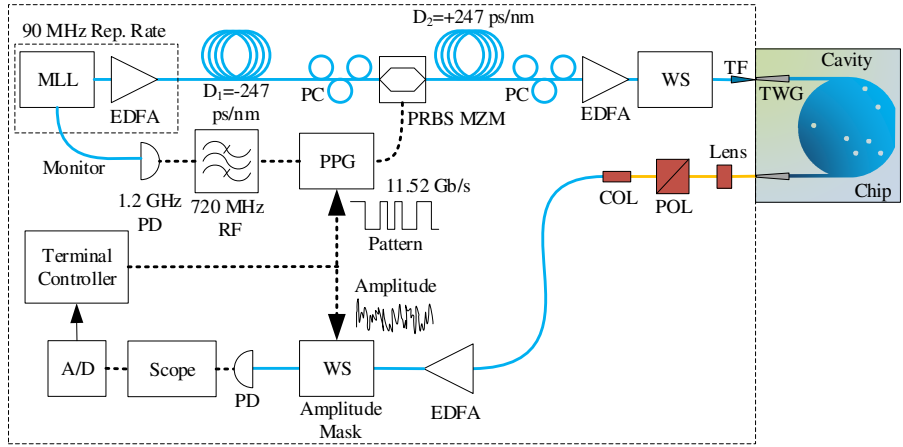

Fig. 2. Silicon photonic PUF experimental setup. A 90 MHz mode-locked laser (MLL) generated ultrashort pulses, i.e. < 300 fs full-width half max (FWHM). These are amplified via an erbium doped fiber amplifier (EDFA) and sent through dispersion compensating fiber (DCF) to spread the pulse to about 11 ns. The pulses are sent through a polarization controller (PC) and then through a Mach Zehnder modulator (MZM) to encode a pseudo-random binary sequence onto each pulse in real time. This is performed by taking the synchronized monitor port of the MLL and first detecting each pulse with a photo-diode (PD). The signal is passed through a low pass filter and then used as a clock signal for the pulse pattern generator (PPG) which is connected to the MZM. The spectrally-modulated pulses are sent through anomalous DCF to compress them back to < 3ps. The pulse train is sent through another PC, EDFA, and a programmable spectral filter or WaveShaper (WS) (for additional dispersion compensation) prior to insertion into a tapered fiber at the chip edge aligned to a tapered waveguide (TWG) for fiber-to-waveguide coupling. The pulses excite many modes within the cavity and each optical response is collected via a focusing lens and collimator (COL) to couple back into a single mode fiber (SMF). A polarizer (POL) is used on the output to select a polarization state under test. The output pulses are amplified for detection by an EDFA and sent through a WS to pass each pulse through a 296-feature spectral pattern over the spectral bandwidth of each pulse. The pulses are detected using a PD and read via an oscilloscope for further post-processing to convert into the final binary key material.

The native 300-fs duration pulses from the mode-locked laser are used with a novel ultrafast pulse shaper that spectrally encodes the amplitude of each individual pulse as follows: [18] dispersion compensating fiber (DCF) stretches the 300-fs mode-locked laser (MLL) pulse to greater than 11 ns. The temporally dispersed spectrum is spectrally amplitude encoded by a length 128 pseudorandom binary sequence (PRBS) at 11.52 Gbit/s that is synchronized to the MLL. There is some overlap between time stretched pulses at this stage and thus neighboring pulses share some temporal features. However, they are mapped to different wavelengths and thus involve different parts of the pattern, i.e. do not spectrally overlap. This allows the patterns on each pulse to remain incoherent while providing more features on each pulse. We achieve 94 features within the 3-dB bandwidth of each pulse.

After spectral patterning, the pulses are compressed using $D_2=+247$ ps/nm fiber to 6 ps. These pulses are amplified with an EDFA (Amonics C-band Erbium Doped Fiber Amplifier AEDFA-PA-35, 150mW) to an average power of 64 mW and coupled into the chip as before. The output pulses are then sent to another EDFA (Amonics C-band Erbium Doped Fiber Amplifier AEDFA-PA-30, 20 mW) to pre-amplify for detection. The amplified pulse is sent through a spectral filter (WaveShaper 1000s, C-Band), which applies 296 amplitude features to the pulse. The input pulse bandwidth (1535-1575 nm) was not perfectly aligned with the spectral filter used in the experiment (1527.4-1567.5 nm), thus some of the spectrally-encoded information was lost. The filtered pulse was sent to a photo-diode and detected with an ADC convertor (Prologic Designs ADC16100LAN with the ability to store over four million samples) and was synchronized to the MLL.

*Post-Processing Algorithm*

The measured power level from the system is transformed into a binary sequence though an algorithm that enhances its indistinguishability from a true random binary sequence. In the oscilloscope detection configuration, the original measurements are interpolated to enhance the available points between peak power samples. The algorithm then finds the peaks of the resultant power signals that correspond to individual challenge pulses facilitated by use of a repeating header sequence. To enhance the SNR of the sampled signals, an optimization routine is run on the sampled data to determine an ideal integration window. Once determined, a window centered at the peak centroid for each response is integrated to produce an integrated power measurement. In the Prologic ADC detection configuration, this step is not required as there is only a single synchronized point per detected pulse. A probability density function (PDF) is calculated for the challenge set. A histogram equalization algorithm is used to calculate non-uniform levels that will make any subsequently collected responses equi-probable when converted to binary. The power samples are then digitized and converted to binary against the non-uniform levels (resampling). An XOR operation is performed on adjacent sequences [19] and a number of MSBs are kept from each sample and appended together to create a single bit sequence representative of the challenge set. The resampling bits and the number of kept MSBs are optimized to minimize error.

*Security Requirements*

In order to exploit this scheme for secure communications, we first define the associated security requirements. The unconditional security for a OTP protocol is guaranteed only if the message is mixed with a random key and never reused [13]. We assume that the output of a cryptographic hash function is sufficient for use as statistically random key material [14] and that the reuse requirement may be met with proper protocol design and execution. The security of this approach does not depend on the security of any electronically stored data as all such data can be made public without loss of security. The primary requirement is that the adversary must not be able to efficiently copy or model the operation of the PUF [3]. Our previous work has demonstrated the unclonability of our photonic micro-cavity PUFs [9], which we verify for the system conditions here and it thereby satisfies this requirement. Further, entropic security, i.e. with an observation of some ciphertext, an adversary will not be able to compute any predicate on the ciphertext with meaningfully larger probability than an adversary who does not possess the ciphertext [15,16], is guaranteed only if the target data has high entropy [17].

## 3. Detector Performance and Key Material Quality

For use of these silicon photonic PUFs in a OTP encrypted communications system it is critical that a large amount of key material can be extracted from a single device. To investigate the extraction of long binary keys we probe a single PUF device with a sequence of $2\times10^7$ uniquely patterned pulses and apply 31 orthogonal Hadamard spectral filter patterns to the output, thus generating $6.2\times10^8$ analog pulse energy samples, which we convert to binary keys through the methods described as follows.

*Robustness and Repeatability Analysis*

A robustness and repeatability analysis was performed on the experimental designs. A common set of 8550 pseudo-randomly generated patterns of 128 bits were used for the pattern set. The Prologic Designs ADC16100LAN, Analog-Digital Converter has the ability to collect over 4 million samples with one sample for each input pattern yielding enough memory for 468 repetitions. All of the experimental data went through the post-processing algorithm to represent the 8550 patterns as a single bit sequence. Inter and intra-parameter FHD were calculated for every possible combination of bit sequences. For each distribution, the mean and standard deviation were also calculated. These values were used to calculate the effective number of independent bits via $N=p(1-p)/\sigma$, where $p$ is the experimental mean and $\sigma$ is the experimental standard deviation [4]. These results were used to fit a binomial distribution to the different configuration parameter sets and consequently calculate probabilities of error correction on the response from an illegitimate device from the corresponding cumulative binomial distributions.

*Entropy and Compressibility*

To assess the quality of this key material we analyze its entropy and compressibility. First we quantify the number of bits that can be accurately extracted from each analog sample by measuring the effective number of bits (ENOB) resulting from the signal to noise ratio of the samples [Fig. 3(a)] and find the mean to be 9.64 bits. We then compute the mean entropy rate of the raw analog samples as a function of the number of sampling bits [Fig. 3(b)] by measuring the probability density function (PDF) at each sampling level. At a sampling value of 10 bits to approximate the ENOB [Fig. 3(c)], we find that the entropy rate is 9 bits, indicating the maximum rate of entropy per sample. However, for key generation it is desirable that each level is equiprobable. Thus, we resample using a nonuniform level spacing and analysis of this operation indicates that 4 or fewer of each sample's most significant bits (MSBs) are sufficiently repeatable for key generation [Fig. 3(d)], resulting in approximately 2.4 Gb of key material.

In order to estimate a lower bound on the binary response entropy, we apply an open-source context tree weighting (CTW) compression scheme [20,21], operating with its default parameters and a tree-depth of 6 to 24 blocks of 100 Mb of key material as demonstrated in [3]. The sample key material is seen to be incompressible [Fig. 3(e)], suggesting a per-bit entropy of exactly one. Thus the silicon photonic PUF device has an information capacity of at least 2.4 Gb. Notably, given the device's size this results in an information density of 2.3 Pb/in$^2$ which is several orders of magnitude beyond industrial memory storage methods, i.e. compact disks (0.9 Gb/in$^2$), digital video discs (2.2 Gb/in$^2$), hard disk drives (1 Tb/in$^2$), and solid-state memory (2.8 Tb/in$^2$) [22–24].

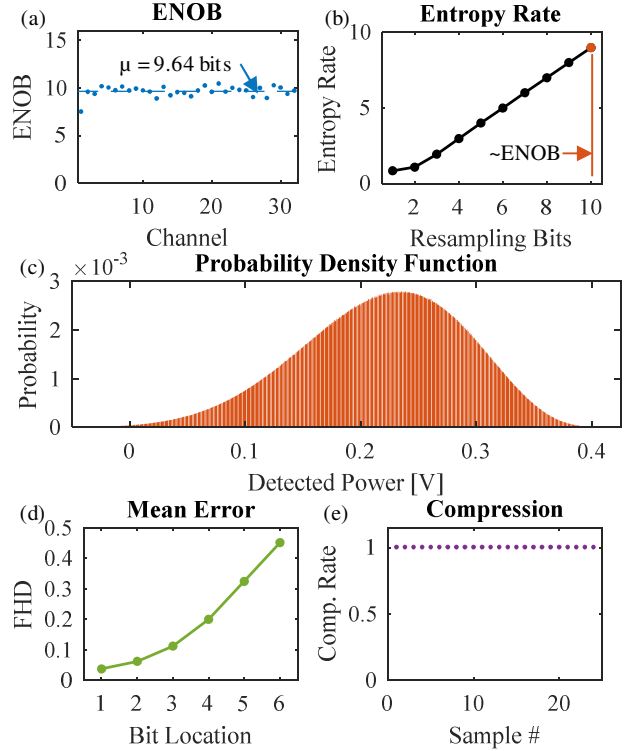

Fig. 3. PUF output evaluation. (a) Effective number of bits (ENOB) for each detection channel formed from a set of order-32 Hadamard orthogonal spectral patterns applied to 20×10$^6$ PUF responses prior to detection. (b) Mean entropy rate across all channels at different resampling levels in post-processing. (c) Probability density function (PDF) of the detected signal across all channels and responses for a resampling to 10 bits in post-processing. (d) Mean error by bit location (big endian) in terms of fractional Hamming distance (FHD) between binary responses generated from successive repetitions of the same challenge to the system. (e) Compression rate of processed binary responses via the CTW algorithm for 24 blocks of 100 Mbits samples of key material. The observed mean compression rate (1.0061) indicates the incompressibility of the PUF responses prior to fuzzy extraction.

## 4. Secure Communications Protocol

Next we investigate the secure communications performance of silicon photonic PUFs by implementing an encrypted OTP communication channel between two parties [13], Alice and Bob, each of whom possesses a distinct photonic PUF [Fig. 4]. In order to communicate, they synchronize their photonic PUFs to generate a shared key by first meeting physically or over a known secure channel. This process also generates public helper data that aids in key recovery and negligibly reduces the security of the system. To send a message, Alice reconstructs her private key, encrypts a message of equal length, and sends the ciphertext over a public communication channel. Bob reconstructs his private key and performs an exclusive-or (XOR) operation with a previously generated shared key to recover Alice's private key with which he recovers Alice's message. Alice and Bob would communicate until their key space is exhausted at which time they generate more shared key material. This communications protocol can be enhanced by requiring authentication [8,9] to access the public dictionary to mitigate denial of service attacks.

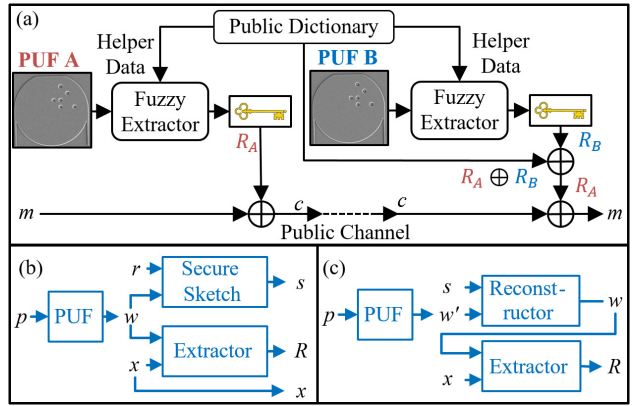

Fig. 4. (a) Encrypted communication protocol. Two endpoints, A and B, have unique photonic PUFs. A fuzzy extractor is applied to the response sequence derived from PUF A to recover a specific block of key material with the help from data in a public dictionary. A message is encrypted via digital exclusive-or (XOR) with that key and is sent through a public channel. Endpoint B recovers their unique key in the same manner as A. The key from A is recovered by B via digital XOR with a previously constructed shared key. This recovered key from A is used to recover the message from the ciphertext via digital XOR. (b) Generation procedure. A challenge $p$ interrogates the PUF resulting in a binary response w, which is sent into the secure sketch (SS) and extractor. The SS takes w and a random value $r$ to generate helper data $s$. The extractor ingests w and a random value $x$ to produce key $R$. Both $s$ and $x$ are stored in a public dictionary as helper data. (c) Reconstruction procedure. Challenge $p$ interrogates the PUF which produces $w'$ which may be different than w given noise. The reconstructor takes helper data $s$ and response $w'$ to reconstruct w which the extractor uses, with helper data $x$, to reproduce key $R$.

This scheme comprises three building blocks [Fig. 4(b) & (c)]. An *extractor* withdraws information of uniform randomness from the photonic PUF response [17]. The *secure sketch* creates public information from its input that does not provide an adversary a significant advantage [17] and that allows the *reconstructor* to fully reconstruct the original input [25] from one that is reasonably close to the original. The cipher is applied to blocks of PUF key material. If the PUF response is not reconstructed to a predetermined error threshold, no portion of the message in that block is recovered. A fuzzy extractor is used in the dictionary setup to generate keys for public storage and in the communications protocol to recover the original key from a noisy key.

*Fuzzy Extraction Algorithm*

During the dictionary setup procedure [Fig. 5(a)], several repetitions of the response sequence from $PUF_A$ of a single input pattern, $p_i$, are averaged together prior to binary post-processing to form an averaged PUF response $\langle w_i \rangle$ corresponding to the $i^{th}$ block, e.g. $\langle w_i \rangle = \langle PUF_A(p_i) \rangle$. Within the secure sketch, a random word, $k_i$, is selected for each block to generate $r_i = BCH_{enc}(k_i)$ [26] and combined with the averaged PUF response via XOR to generate the helper data $s_i$, e.g. $s_i = \langle w_i \rangle \oplus r_i = \langle PUF_A(p_i) \rangle \oplus BCH_{enc}(k_i)$. Within the fuzzy extractor, the averaged PUF response is combined with a randomly generated seed, $x_i$, via XOR and then input into the SHA-256 hash function to generate $\langle R_i \rangle = SHA(\langle w_i \rangle \oplus x_i) = SHA(\langle PUF_A(p_i) \rangle \oplus x_i)$. The random seed is stored as helper data in the public dictionary. Likewise, the same process follows for the second PUF such that $\langle R_j \rangle = SHA(\langle w_j \rangle \oplus x_j)$ with $s_j$ and $x_j$ being stored as helper data in the public dictionary where such data is stored by the $j^{th}$ block corresponding to $PUF_B$. In addition to the helper data, the input patterns, $p_i$ and $p_j$, are also stored in the dictionary. Lastly, a shared key corresponding to both PUFs is calculated by combining $\langle R_i \rangle$ and $\langle R_j \rangle$ via XOR, e.g. $k_{ij} = \langle R_i \rangle \oplus \langle R_j \rangle$, and storing in the public dictionary.

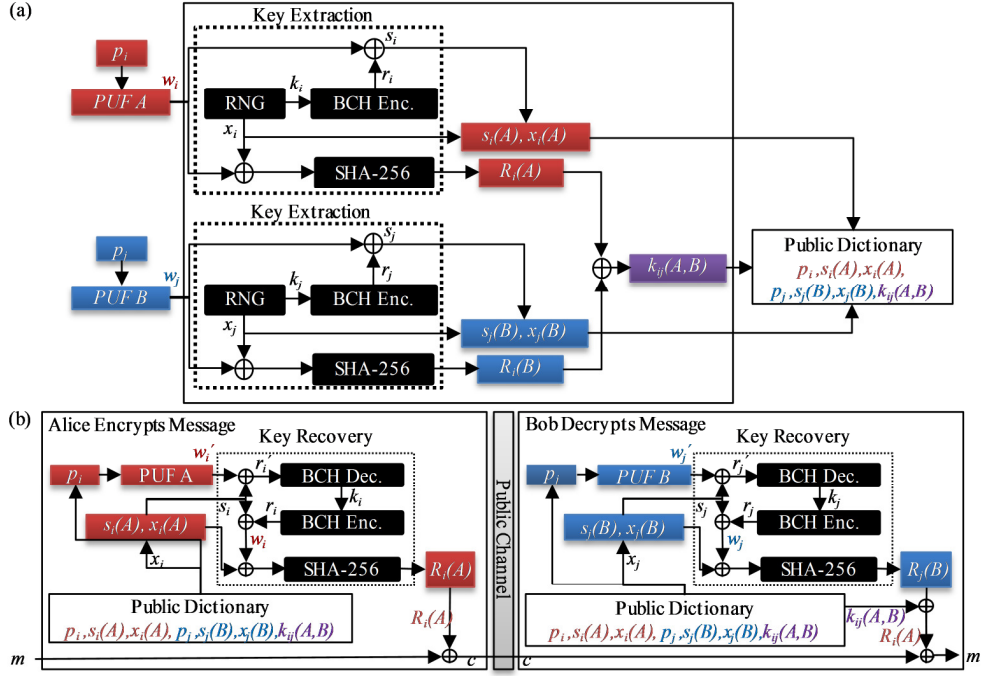

Fig. 5. Fuzzy Extraction Dictionary Setup and Communications Protocol. (a) Dictionary Setup.
(b) Communications Protocol.

In order to encrypt a message [Fig. 5(b)], Alice will recover her key by first querying her token and recording a single shot response from her PUF, e.g. $w_i = \text{PUF}_A(p_i)$, then combining the response with the helper data, $s_i$, to recover some corrupted $r'_i = w_i \oplus s_i$ from which $k_i$ is perfectly recovered via BCH decoding, so long as $e_i = w_i \oplus \langle w_i \rangle \leq (d_{min} - 1)/2$ where $d_{min}$ is the minimum distance of the BCH code. $k_i$ is encoded with the same BCH code to fully recover $r_i$ from $r'_i$. The errors in the PUF response are corrected and the averaged response is recovered by combining $s_i$ and $r_i$ via XOR to achieve $\langle w_i \rangle = s_i \oplus r_i$. If the Hamming distance between $w_i$ and $\langle w_i \rangle$ is greater than $(d_{min} - 1)/2$, then either a different BCH codeword altogether will result (called "decoder error"), or the BCH decoder will be unable to output a binary word at all ("decoder failure"). The recovered response is XORed with the stored random seed $x_i$ and hashed to form $R'_i = \text{SHA}(w'_i \oplus x_i)$. Any differences between $w'_i$ and $\langle w_i \rangle$ will be amplified via the hash function as the hashes will be uncorrelated [14]. Lastly, the message, $m$, is combined via XOR with $R'_i$ to form the ciphertext, $c = m \oplus R'_i$.

In order to decrypt the ciphertext, Bob follows the same recovery process as Alice to recover his key, $R'_j = \text{SHA}(w'_j \oplus x_j)$. Bob then combines via XOR his recovered key, $R'_j$, with the shared key, $k_{ij}$, to recover a variant of Alice's key, e.g. $R''_i = k_{ij} \oplus R'_j = \langle R_i \rangle \oplus \langle R_j \rangle \oplus R'_j$. Bob then combines Alice's recovered key via XOR with the ciphertext to recover the message, e.g. $m' = c \oplus R''_i = m \oplus R'_i \oplus \langle R_i \rangle \oplus \langle R_j \rangle \oplus R'_j$. Variations in laser and modulator outputs, power loss, vibration, and detector noise contribute to overall system noise which unavoidably corrupts the PUF response, causing the same patterned pulse to produce varying responses. As such, the probability of an error in the recovered message is based on the probability that each PUF response will deviate from the expected response beyond the error correction capability of the error correction code (ECC). The message is recovered without error, $m' = m$, if and only if $w_i \cong \langle w_i \rangle$ and $w_j \cong \langle w_j \rangle$, i.e. if $e_i$ and $e_j$ are correctable by the BCH code used.

The probability of an error in the perfect recovery of the hashes for $PUF_A$ is $Pr_A(e) = Pr(R_i' \neq \langle R_i \rangle)$. This is equivalent to the probability that a PUF response has more errors with respect to the averaged PUF response than the error correction threshold, e.g. $Pr_A(e) = Pr(FHD(w_i', \langle w_i \rangle) \geq \alpha)$, where $\alpha$ is the threshold at which BCH can correct all of the errors in the response. This probability is equal to the complementary cumulative binomial probability distribution evaluated at that threshold. Likewise, for $PUF_B$, the probability of an error is $Pr_B(e) = Pr(FHD(w_j', \langle w_j \rangle) \geq \alpha)$. Therefore, the overall probability that a block of data is corrupted is the likelihood that either of the PUF responses are not fully recovered, e.g. $Pr(m' \neq m) = 1 - [(1 - Pr_A(e)) \times (1 - Pr_B(e))]$.

*Experimental Results*

To quantify the robustness of this approach, we interrogate all of the photonic PUFs with an identical sequence of 8550 uniquely patterned ultrafast optical pulses. The optical response sequence from each cavity is passed through a pseudorandom spectral mask with 296 features across the optical bandwidth and the transmitted analog pulse energy sequence is recorded using a photodetector and ADC and converted into a binary key as before. In this case, we save the three MSBs of the key material derived from each pulse, which are concatenated to form a 25,650-bit binary response for each device. The binary response is decomposed into 100 blocks of 255 bits and inserted into a fuzzy extractor, which produces helper data and a binary hash sequence to generate suitable keys for secure communications. To best represent a real-world implementation, the measured analog pulse energy sequence from 460 repetitions is averaged together prior to binary conversion for generation of the shared key, whereas for message encryption a single-shot (not averaged) sequence is employed. Thus for encryption the key material is extracted from the PUF at 0.27 Gbps.

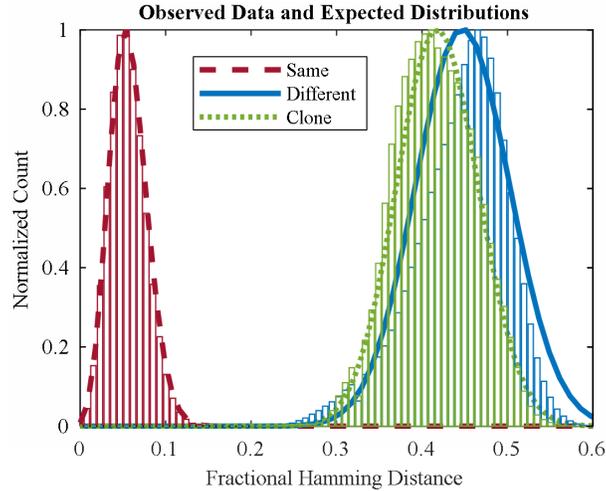

Fig. 6. Normalized FHD binomial distributions and histograms for binary responses from same, different, and cloned designs, resampling to 3 bits post-analog-to-digital conversion (ADC) at a block size of 256 bits. All responses were generated from the quasi-Transverse Electric (TE) polarization state.

To analyze the uniqueness and repeatability of the generated binary keys the fraction of positions in which the binary sequences differ (the "fractional Hamming distance" or FHD) is calculated between each individually generated key and the averaged keys used to generate the dictionary for all 22 devices. The histogram of FHDs between each individual binary key from each device and the averaged binary key from that same device forms the "same" distribution [Fig. 6]. The histogram of FHDs between each individual binary key from each device and the averaged binary key from each different device forms the "different" distribution. Likewise, the histogram of FHDs between each individual binary key from each device and the averaged binary key from each respective cloned device forms the "clone" distribution. The clear separation of these histograms demonstrates the distinguishability of binary keys from different devices including exact clones and the small FHD of the "same" distribution demonstrates the robustness of extracting the binary key material at high speed from the physical device. A binomial probability mass function is fit to each histogram and is used to estimate the probability that a given PUF response will be beyond the error correction capabilities of the error correcting code (ECC) at a selected code rate, thus corrupting an entire block of data. The probability that a device of different design or its clone [9] could respond within the correctable range of the code is negligible across all possible code rates indicating strong unclonability.

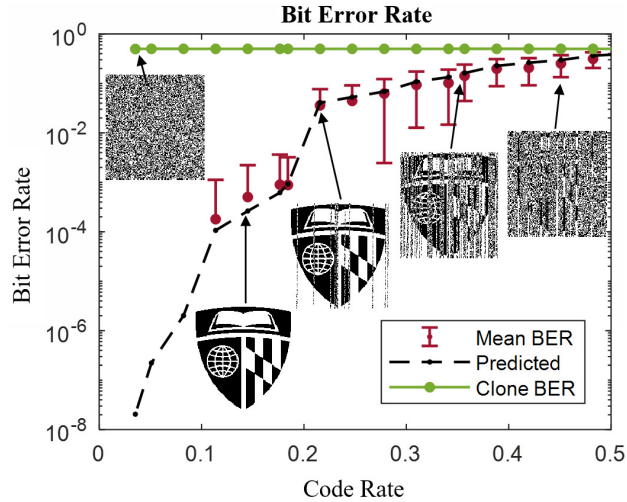

Fig. 7. The mean experimental and predicted BER are shown for 57 combinations of different devices communicating a test image of 25,575 bits at various BCH code rates. The upper and lower bars indicate a two standard deviation bound relative to the mean BER. Lower bounds for low code rates are not shown due to perfect message reconstruction. The inset images show the recovered message (university seal) at various code rates. The mean experimental BER vs code rate for an adversary attempting to decrypt a message using the intended endpoint's clone are shown for the same 57 combinations, demonstrating unclonability. The lowest observed clone BER is 0.483.

To characterize the system BER, 57 different pairs of unique photonic PUFs were used to communicate a message of 25,575 bits. The BER at different BCH code rates is shown in Fig. 7. Communication with a mean BER $< 10^{-5}$ is observed at code rates of $< 0.1$. Notably, at equivalent code rates (0.035) to previous approaches [3], we achieve an eight order of magnitude improvement in BER.

## 5. Security Analysis

One approach to eavesdrop or maliciously send messages is for an adversary to obtain the cavity design and attempt to fabricate a clone of the device. However, the ray chaotic cavity design makes a device's behavior highly sensitive to structural variations arising from the fabrication process such as sidewall roughness, film thickness, resist granularity, and material impurities. To directly investigate this attack, we fabricated two such clones of each device in close proximity on the same chip, thus maximizing consistency of fabrication conditions across devices, and compared the keys generated by legitimate and cloned devices. The BER using cloned devices is shown in Fig. 7. Notably, at a code rate of 0.1 where we achieve error free communication of our 25,575-bit message with a legitimate device, the probability of a successful decryption with a cloned device is on the order of $10^{-11}$, demonstrating the near impossibility of using a cloned device to eavesdrop or maliciously communicate.

Beyond cloning, an adversary could capture a device and attempt to fully characterize it. However, the nonlinearity of our devices protects against such an attack by greatly increasing the amount of information that an attacker would need to characterize [9]. Specifically, in a linear system, the mapping from the system input to its output may be represented as a linear combination of its input symbols in the form of a transmission matrix whose uniqueness is limited by the number of orthogonal input vectors. If an adversary can observe and characterize this transmission matrix, and compute its inverse or pseudoinverse, they can obtain, exactly or approximately, any input given the output and vice-versa. In contrast, in a nonlinear system, the transmission function is a system of nonlinear relationships for which no such inversion exists. As the PUFs investigated here are nonlinear devices, this characterization would potentially require probing of up to $2^{128}$ different challenges and recording the subsequent responses. With our 90-MHz repetition rate laser source, that characterization would take up to $10^{23}$ years and would push the adversary towards other attacks. For example, the adversary could use the input pattern data stored in the public dictionary to reduce their search space. Provided the transport time is carefully monitored, a suitable number of patterns occupy the public dictionary, and patterns are chosen at random, this attack is sufficiently mitigated. Further, if access to the public dictionary is protected using a PUF authentication approach [9] or equivalent public key architecture, then the adversary will have no advantage. Should the adversary record the encrypted communications channel and steal a device at a later date, they could attempt to decrypt previously sent messages. Protocol level enhancements applying ephemeral session keys could mitigate this attack.

While we used the BCH code and the SHA-256 hash in this demonstration, any suitable code and hash can be used. An adversary wishing to reverse the hash function would need to do so for each block; we assume this is sufficiently difficult [27,28] for the adversary to avoid this attack altogether. In order to verify the independence of generated keys, we calculate the entropy of 1000 privacy amplified keys as $(\mu*(1-\mu))/\sigma^2$, where $\mu$ is the mean and $\sigma$ is the standard deviation of the FHD distribution yielding 256.1 bits [4]. Based on this evaluation, the output strings of the fuzzy extractor contain full entropy [29], as their length is 256 bits. Thus, the probability of guessing the true PUF response from the privacy amplified key would be $2^{-256}$, i.e., an information theoretic security of 256 bits.

## 6. Summary, Discussion, and Future Work

In conclusion, we demonstrate encrypted communications using key information extracted from information-dense nonlinear silicon photonic PUFs. Compared to previous work on optical scattering based PUFs [3], this novel photonic PUF provides orders of magnitude improvement on channel BER and reductions in physical size while providing full compatibility with integrated circuits and telecommunications systems. The device also provides information densities far beyond current storage media. Given the growth of the size of the public dictionary, this method may find best application to the exchange of secure keys for modern cryptographic ciphers. We expect that the approach can scale to support the demands of high-speed transmission through use of higher repetition rate and bandwidth laser sources, spectro-temporal multi-level amplitude and phase encoding, as well as temporal multiplexing to enhance key generation rates.

**Funding**

The Johns Hopkins University Catalyst Award Program.

**Acknowledgments**